\documentclass[12pt]{iopart}

\usepackage{iopams}

\def\I{{\mathbb{I}}}
\begin{document}

\title{Massive gravity from bimetric gravity}

\author{Valentina Baccetti, Prado Mart\'{\i}n-Moruno,\\ and Matt Visser}

\address{School of Mathematics, Statistics, and Operations Research
Victoria University of Wellington 
PO Box 600, Wellington 6140,
New Zealand}
\ead{valentina.baccetti@msor.vuw.ac.nz, prado@msor.vuw.ac.nz and matt.visser@msor.vuw.ac.nz}
\begin{abstract}
We discuss the subtle relationship between massive gravity and bimetric gravity, focusing particularly on the manner in which 
massive gravity may be viewed as a suitable limit of bimetric gravity.
The limiting procedure is more delicate than currently appreciated. Specifically,  this limiting procedure should not 
unnecessarily constrain the background
metric, which must be externally specified by the theory of massive gravity itself.
The fact that in bimetric theories one always has two sets
of metric equations of motion continues to have an effect even in the massive gravity limit, leading to additional
constraints besides the one set of equations of motion naively expected.
Thus, since solutions of bimetric gravity in the limit of vanishing kinetic term are also solutions of massive gravity,
but the contrary statement is not necessarily true, there is not complete continuity in the parameter space of the theory.
In particular, we study the massive cosmological solutions which are continuous in the parameter space, showing that many interesting
cosmologies belong to this class.

\end{abstract}

\vspace{2pc}

\maketitle

\section{Introduction}\label{uno}

Massive gravity has recently undergone a significant surge of renewed interest. Since de~Rham and Gabadadze~\cite{deRham:2010ik}, (see also 
de~Rham, Gabadadze, 
and Tolley~\cite{deRham:2010kj}), demonstrated that it is possible to develop an extension of the Fierz--Pauli mass 
term for linearized gravity~\cite{Fierz:1939ix},  one
that avoids the appearance of the Boulware--Deser ghost~\cite{Boulware:1973my}, at least up to fourth order in non-linearities, 
activity on this topic has become intense --- with over 50 articles appearing in the last two years. 

Subsequently, both Hassan and Rosen~\cite{Hassan:2011vm}, and Koyama, Niz, and Tasinato~\cite{Koyama:2011yg},  have
independently re-expressed and re-derived the theory of de~Rham and Gabadadze --- simplifying the treatment and shedding additional light on its characteristics.
In fact, Hassan and Rosen took a significant step further by extending the theory to a general background metric~\cite{Hassan:2011vm}.
Rapidly thereafter,  several papers appeared studying the foundations of this 
theory~\cite{Koyama:2011zz,Chamseddine:2011mu,deRham:2011rn,deRham:2011qq,Kluson:2011qe,Golovnev:2011aa},
and proving the absence of ghosts in the nonlinear theory~\cite{Hassan:2011hr, Hassan:2011tf, Hassan:2011ea,Hassan:2012qv}.
In particular, Hassan and Rosen showed that the introduction of a kinetic term for the background metric in the ghost-free massive gravity
leads to a bimetric gravity theory which is also ghost-free~\cite{Hassan:2011zd}. It should be noted that in this case
the background metric is not only an externally-specified kinematic quantity, but also has its own dynamics, 
acquiring the same physical status as  the ``foreground'' metric~\cite{Isham:1971gm}.
More recently, ghost-free multi-metric theories have also been considered~\cite{Hinterbichler:2012cn,Hassan:2012wt}.

Any cautious and physically compelling approach to massive gravity should not only respect the beauty of standard
general relativity,  but also enhance it in some manner, by embedding standard general relativity in some wider parameter space. 
Physically, one might hopefully expect that the observational predictions of the extended theory should be continuous in these extra parameters, and that general relativity
would be recovered by taking the limit for a zero graviton mass in massive gravity theories.
However, this is not necessarily the case, since (as is well known) the predictions of massive gravity often qualitatively differ from those of general relativity
even when the graviton mass vanishes. This effect is known as the van~Dam--Veltman--Zakharov (vDVZ) discontinuity~\cite{vanDam:1970vg, Zakharov:1970cc}.
The Vainshtein mechanism provides us (under some appropriate conditions) with a way to avoid (or rather ameliorate) 
such a discontinuity~\cite{Vainshtein:1972sx} (see also~\cite{Damour:2002gp,Babichev:2009jt,Babichev:2009us,Babichev:2010jd} and~\cite{Koyama:2011yg}). 
Furthermore, as has been pointed out in reference~\cite{Hassan:2011vm}, the vDVZ discontinuity can sometimes be arranged to be absent when one considers the background metric to be non-flat in massive gravity, which is certainly the more general situation. That is, it seems that the predictions of massive gravity might be arranged to be continuous in
parameter space when suitably taking into account a curved background~\cite{Rubakov:2008nh,  Hinterbichler:2011tt} --- this is one of several reasons for being interested in non-trivial backgrounds. 

But why, in the first place, should one modify general relativity in such a manner and even entertain the possibility of massive gravity?
References~\cite{Rubakov:2008nh} and~\cite{Hinterbichler:2011tt} provide historical reviews of the motivations.
On the other hand, over the last decade the theoretical revolution based on the inferred accelerated expansion of the universe has become a good reason for considering a wide variety of possible
modifications of general relativity~\cite{Nojiri:2010wj, Sotiriou:2008rp, Sotiriou:2006qn, Capozziello:2007ec, Faraoni:1998qx}.  
Consequently, when de Rham, Gabadadze, and Tolley found a family of
ghost-free (flat-background) massive gravity theories~\cite{deRham:2010ik,deRham:2010kj}, their cosmological consequences were quickly 
analyzed~\cite{deRham:2011by, Chamseddine:2011bu, D'Amico:2011jj, Gumrukcuoglu:2011ew},
even though the theory had not yet been formulated to be compatible with a non-flat background metric.  
Although these studies can be considered as a first attempt to understand the cosmological consequences of ghost-free massive gravity, 
their results are severely limited by the (with hindsight unnecessary, and perhaps even physically inappropriate) 
choice of a flat background metric. 
Indeed, assuming a flat background is not the most general situation,  and it is even contra-indicated 
when considering
black hole geometries~\cite{Deffayet:2011rh} or cosmological scenarios~\cite{Visser:1997hd}.
Later on, independently and almost simultaneously, three groups have considered cosmological solutions in bimetric 
gravity~\cite{Volkov:2011an, vonStrauss:2011mq, Comelli:2011zm}, 
while also studying massive-gravity cosmologies more or less in passing.
However, as we will show in this paper, considering solutions of massive gravity as a limit of 
bigravity theories can be much more subtle than expected.

At this point one might still reasonably wonder whether bimetric gravity is itself well-behaved in parameter space. 
Although a background kinematic metric has provided us with a continuous limit of massive gravity with respect 
to general relativity, it could still be
that the introduction of a kinetic term for the background metric, (with its associated dynamics), disrupts 
coherence with respect to massive gravity. In other words,
it may well be that the physical predictions of bimetric gravity differ from those of massive gravity in the
limit where this theory should be recovered.
To settle this issue, in this paper we study the massive gravity limit of bimetric gravity in full generality. As we will show, the solutions of
bimetric gravity in the limit where the kinetic term for the background metric vanishes will also be solutions of massive gravity compatible
with a non-flat background metric, \emph{but not necessarily vice versa}.

This paper is organized as follows:
In section~\ref{dos} we (briefly) summarize some previous results of massive gravity and bimetric gravity. 
In section~\ref{tres} we explore various limits by which bimetric gravity might be used to reproduce massive gravity. 
In particular we investigate how the vanishing kinetic term limit should be considered, see section~\ref{sec31}, concluding that
the limit procedure is not implying the need for a flat background metric in massive gravity.
Consideration of this limit allows us to obtain the cosmological solutions which are continuous in the parameter space
of massive gravity, see section~\ref{sec32}. We study these solutions which can be classified as general solutions, continuous cosmological
solutions of any ghost-free theory, and special case solutions. Although the second group are solutions only for a particular kind of
massive gravity models, we show that they present some features of particular interest.
In section~\ref{sec33} we briefly comment the consequences of retaining some effects of the background matter 
when taking this limit.
Section~\ref{cuatro} contains some discussion and 
conclusions, while some purely technical formulae and computations are relegated to  \ref{appendix-a}, \ref{appendix-b}, 
and \ref{appendix-c}.

\section{Massive gravity and bimetric gravity}\label{dos}

As is already rather well-known~\cite{Boulware:1973my} (see also~\cite{Visser:1997hd}),  
in order to consider massive gravity one is forced to introduce a new rank-two tensor $f_{\mu\nu}$ whose kinematics is at this stage externally specified,  
and not governed by the theory itself.  This new  rank-two tensor $f_{\mu\nu}$ can best be interpreted as a ``background'' metric, 
not necessarily flat, with linearized fluctuations $h_{\mu\nu}$ of the physical ``foreground'' metric $g_{\mu\nu} = f_{\mu\nu} + h_{\mu\nu}$ 
satisfying a massive Klein--Gordon equation~\cite{Boulware:1973my} (see also~\cite{Visser:1997hd} and~\cite{ArkaniHamed:2002sp}). 
That is, the (full non-linear) mass term appears in the action as some interaction term which depends algebraically on the tensors 
$f_{\mu\nu}$  and $g_{\mu\nu}$, through the quantity 
\begin{equation}
(g^{-1} f)^\mu{}_\nu = g^{\mu\rho}f_{\rho\nu},
\end{equation}
and which in the linearized limit reproduces a suitable quadratic mass term.  On the one hand, there is a non-denumerable infinity of 
such interaction Lagrangians, on the other hand, almost all of them lead to the physically unacceptable 
Boulware--Deser ghost~\cite{Boulware:1973my}.
For example, a particular specification of the interaction Lagrangian given in reference~\cite{Visser:1997hd}
led to the cosmological models presented in 
references~\cite{Alves:2007hr, Alves:2009sh, Alves:2010zz, Basilakos:2011fy}, 
which assumes a flat background metric contraindicated for this situation, and
reference~\cite{deRoany:2011rk}, which additionally contains specific technical criticisms 
to~\cite{Alves:2007hr, Alves:2009sh, Alves:2010zz, Basilakos:2011fy}.
Different approaches include the Lorentz-violating massive 
gravity~\cite{Dubovsky:2004sg,Bebronne:2007qh,Bebronne:2009mz} and the 3-dimensional 
``new massive gravity''~\cite{Bergshoeff:2009hq,Bergshoeff:2009aq}.
The novelty of the last two years is that the interaction Lagrangian has now been very tightly constrained and fixed to depend only on a
finite number of parameters~\cite{Hassan:2011vm} by requiring the theory to be ghost-free. 
A quite unexpected technical result (from the  Boulware--Deser point of view~\cite{Boulware:1973my}, see also~\cite{Visser:1997hd}) is that the dependence of this ghost-free interaction term on the background and foreground metrics is through the square-root quantity 
\begin{equation}
\gamma^{\mu}{}_{\nu}=\left(\sqrt{g^{-1}f}\right)^{\mu}{}_{\nu},\,\, {\rm that\, is}\,\,\,\,  \gamma^{\mu}{}_{\sigma}\gamma^{\sigma}{}_{\nu}=g^{\mu\sigma}f_{\sigma\nu}.
\end{equation}

On the other hand, bimetric gravity was first introduced by Isham, Salam and Strathdee~\cite{Isham:1971gm} (see 
also~\cite{Aragone:1972fn, Salam:1976as, Isham:1977rj}) to account for some features of strong interactions, and it was later 
rejuvenated by Damour and Kogan in order to address new physics scenarios~\cite{Damour:2002ws} 
(see also~\cite{Damour:2002wu, Blas:2005yk,Berezhiani:2008nr}).
It has also been recently proven to be ghost-free when considering the same interaction term as 
for massive gravity~\cite{Hassan:2011zd}.

Let us now focus our attention on 4-dimensional (and Lorentz invariant) massive gravity.
We can express the action generally as
\begin{equation}\label{actionmg}
S_{MG}=-\frac{1}{16\pi G}\int d^4x\sqrt{-g} \left\{R(g)+2\,\Lambda-2\,m^2 \, L_{{\rm int}}(g^{-1} f)\right\}+S_{({\rm m})},
\end{equation}
with $S_{({\rm m})}$ describing the usual matter action, with matter fields coupled only to the foreground metric $g_{\mu\nu}$,
(and the measure $\sqrt{-g}$\,), to agree with the Einstein equivalence principle. 
The parameter $m$ sets the scale for the graviton mass, and the interaction term $L_{{\rm int}}(g^{-1} f)$ 
is a scalar chosen to be dimensionless.

Regarding bimetric gravity, in addition to the kinetic term of the background metric, 
one must also consider the possibility of a background cosmological 
constant $\bar\Lambda$, and ``background matter'' $\bar S_{({\rm m})}$ coupling to $f_{\mu\nu}$. This now leads to
\begin{eqnarray}\label{actionbg}
S_{BG}&=&-\frac{1}{16\pi G}\int d^4x\sqrt{-g} \left\{R(g) + 2\,\Lambda-2\,m^2 L_\mathrm{int}\right\} +S_{({\rm m})}
\nonumber\\
&&
-\frac{\kappa}{16\pi G}\int d^4x\sqrt{-f} \left\{\mathcal{R} (f)  +2\, \bar\Lambda \right\} +\epsilon\,\bar S_{({\rm m})}.
\end{eqnarray}
The effective Newton constant for the background spacetime/matter interaction is $\epsilon G/\kappa$. The two parameters $\kappa$ and 
$\epsilon$ can in principle be adjusted independently.

The interaction term of the ghost-free theories can without loss of generality be written as~\cite{Hassan:2011vm,Hassan:2011zd} 
(see also the discussion of~\ref{appendix-b})
\begin{equation}\label{int}
 L_{{\rm int}}=e_2(K)-c_3\,e_3(K)-c_4\,e_4(K),
\end{equation}
with
\begin{equation}\label{K}
 K^{\mu}{}_{\nu}=\delta^{\mu}{}_{\nu}-\gamma^{\mu}{}_{\nu},
\end{equation}
and the polynomials $e_i$ (see~\ref{appendix-a} for a more formal definition and properties) are
\numparts
\begin{eqnarray}
e_2(K)&=\frac{1}{2}\left([K]^2-[K^2]\right);\\
e_3(K)&=\frac{1}{6}\left([K]^3-3[K][K^2]+2[K^3]\right);\\
e_4(K)&=\frac{1}{24}\left([K]^4-6[K^2][K]^2+3[K^2]^2+8[K][K^3]-6[K^4]\right);
\end{eqnarray}
\endnumparts
where $[K]=\tr(K^\mu{}_\nu)$, and our definition of $K$, equation~(\ref{K}), agrees with that of 
references~\cite{Koyama:2011yg} and \cite{Volkov:2011an}, but differs from that of reference~\cite{Hassan:2011vm}.
One can now consider the foreground tetrad $e^{\mu}{}_A$, and the background inverse tetrad $w_{\mu}{}^A$ 
defined by~\cite{Volkov:2011an}  
(see also references~\cite{Nibbelink:2006sz, Chamseddine:2011bu, Chamseddine:2011mu, Hinterbichler:2012cn})
\begin{equation}\label{tetrads}
 g^{\mu\nu}=\eta^{AB}\,e^{\mu}{}_A\,e^{\nu}{}_B,\qquad\qquad  f_{\mu\nu}=\eta_{AB}\,w_{\mu}{}^A\,w_{\nu}{}^B,
\end{equation}
where no direct analogy between these expressions and the definition of the St\"uckelberg 
fields should necessarily be deduced
(the relation between quantities defined in the tangent space and in the tetrad basis cannot be thought of as 
in any way recovering any gauge freedom).
This formalism allows one to write the term involving the square root as
\begin{equation}
 \gamma^{\mu}{}_{\nu}=e^{\mu}{}_A\,w^A{}_{\nu},
\end{equation}
by requiring the consistency condition 
\begin{equation}\label{condt}
  e^{\mu}{}_A\;w_{B\mu}=e^{\mu}{}_B\;w_{A\mu},
\end{equation}
which leaves the equations of motion unchanged~\cite{Volkov:2011an}.

The equations of motion for massive gravity are obtained
by varying the action~(\ref{actionmg}) with respect to $g^{\mu\nu}$. These are
\begin{equation}\label{motiong}
 G^{\mu}{}_{\nu}-\Lambda\,\delta^{\mu}{}_{\nu}=m^2\,T^{({\rm eff})\mu}{}_{\nu}+8\pi G \;T^{({\rm m})\mu}{}_{\nu} ,
\end{equation}
with $T^{({\rm m})\mu}{}_{\nu}$ denoting the usual stress-energy tensor of the matter fields, while
\begin{equation}\label{T}
 T^{({\rm eff})\mu}{}_{\nu}=\tau^{\mu}{}_{\nu}-\delta^{\mu}{}_{\nu}\,L_{{\rm int}},
\end{equation}
is the dimensionless graviton-mass-induced contribution to the stress-energy.
It can be noted that $\tau^{\mu}{}_{\nu}$ can be written as
\begin{equation}\label{taudef}
 \tau^{\mu}{}_{\nu}=\gamma^\mu{}_\rho\,\frac{\partial L_{{\rm int}}}{\partial \gamma^{\nu}{}_{\rho}}
 =e^{\mu}{}_B\,\frac{\partial L_{{\rm int}}}{\partial e^{\nu}{}_{B}},
\end{equation}
for any interaction Lagrangian which depends on the metrics through terms of the form $[K^n]$, with $n$ an integer, (or
equivalently $[\gamma^n]$).
In addition, the Einstein tensor of equation~(\ref{motiong}) needs to satisfy the (contracted) Bianchi identity. 
Taking into account the invariance of $S_{({\rm m})}$ under diffeomorphisms, which implies
$\nabla_{\mu}T^{({\rm m})\mu}{}_{\nu}=0$, that identity leads to a constraint on the graviton-mass-induced effective stress-energy:
\begin{equation}\label{Bianchi}
\nabla_{\mu}T^{({\rm eff})\mu}{}_{\nu}=0.
\end{equation}

On the other hand, the bimetric gravity theory now has two sets of equations of motion, 
obtained by varying with respect to the two metrics. 
Thus, in addition to equation~(\ref{motiong}) one has 
\begin{equation}\label{motionf}
 \kappa\,\left(\mathcal{G}^{\mu}{}_{\nu}-\bar\Lambda\,\delta^\mu{}_\nu\right) =
m^2\, \mathcal{T}^{\mu}{}_{\nu} +\epsilon\,8\pi G \,\bar T^{({\rm m})\mu}{}_{\nu} ,
\end{equation}
where the effective stress-energy tensor for the background metric is~\cite{Volkov:2011an,Comelli:2011zm}
\begin{equation}\label{Tf}
 \mathcal{T}^{\mu}{}_{\nu}=-\frac{\sqrt{-g}}{\sqrt{-f}}\;\tau^{\mu}{}_{\nu}.
\end{equation}
Furthermore, the Bianchi-inspired constraint which follows from equation~(\ref{motionf}) is equivalent to
that already obtained in equation~(\ref{Bianchi}), see~\cite{Volkov:2011an,vonStrauss:2011mq}.

It should be emphasized that the equations of motion (\ref{motiong}), (\ref{motionf}) the definition of the 
effective energy-momentum tensor, equations~(\ref{T}), (\ref{taudef}), (\ref{Tf})
and the constraint (\ref{Bianchi}), are all completely independent of the particular form of the
interaction term, this being an automatic result of the fact that the interaction term depends only  on the quantity $g^{-1}f$ through
terms of the form $[\gamma^n]$. See for instance reference~\cite{Visser:1997hd}.
({\bf Warning:} There is a significant typo in reference~\cite{Visser:1997hd}, amounting to accidentally dropping the scalar factor 
$\sqrt{f/g}$ in the effective stress-energy due to the graviton mass term --- see reference~\cite{deRoany:2011rk}  --- fortunately this 
does not quantitatively affect weak-field physics, nor does it \emph{qualitatively} affect strong-field physics --- though it will 
certainly significantly change many of the details.)

If we now consider the specific family of ghost-free theories given by (\ref{int}), we can explicitly express
$\tau^{\mu}{}_{\nu}$ in terms of matrices as
\begin{eqnarray}\label{tau}
\fl \tau=([\gamma]-3)\gamma-\gamma^2+c_3\left(e_2(K)\gamma-e_1(K)\gamma\cdot K+\gamma\cdot K^2\right)
 \nonumber\\
+ c_4\left(e_3(K)\gamma-e_2(K)\gamma\cdot K+e_1(K)\gamma\cdot K^2-\gamma\cdot K^3\right).
\end{eqnarray}
An equivalent index-based formula can be found in reference \cite{Volkov:2011an}.

\section{Continuity of massive gravity  with respect to bimetric gravity}\label{tres}

An implication of the discussion in the previous section is that
if one considers some particular solutions of a bimetric gravity theory,
then those solutions will be more constrained than those corresponding to a massive gravity theory with the same interaction
term. That is because when one considers $f_{\mu\nu}$ to be non-dynamical  (being externally specified by the definition of the theory), the kinetic term for this metric is no longer
present and one should not consider
the variation of the action with respect to this metric. 

Following this spirit, von Strauss \emph{et al}.~have  considered in reference~\cite{vonStrauss:2011mq}
cosmological solutions for the ghost-free bimetric and massive gravity theories, given by the actions~(\ref{actionbg})
and (\ref{actionmg}), respectively, and the interaction term~(\ref{int}).
They have noted consistently that in massive gravity one loses the equations of motion given by equation~(\ref{motionf}).
Nevertheless, in references~\cite{Volkov:2011an,Comelli:2011zm} 
the authors also study cosmological solutions of massive gravity but considering it
as a particular limit of bigravity theory.
They have obtained different and non-equivalent results to that presented in~\cite{vonStrauss:2011mq}
because when considering massive gravity as a limit of bimetric gravity, 
one has not only the equations of motion of
the foreground metric but also additional constraints.
Thus, one may reasonably wonder whether the consideration of this limit implies that there is some
kind of physical discontinuity in the parameter space of massive gravity.

On the other hand, if one wants to consider massive gravity as a limit of bimetric gravity, then this limit
should be carefully taken to avoid inconsistent results.
In several references, presenting both pre-ghost-free and ghost-free analyses,  the authors have concluded or 
implied that the background 
metric should be flat (Riemann-flat), an interpretation that can hide some problems of the theory,
since a flat metric is an incompatible background for the cosmological scenarios that they have studied.
The recent models for ghost-free massive gravity considered in references~\cite{Volkov:2011an, Comelli:2011zm} amount in the current language 
(and now including the possibility of a background cosmological constant) to taking 
the limit $\kappa\to\infty$ while holding $m$ fixed and setting $\epsilon=0$. The background equations of motion then degenerate to 
$\mathcal{G}_{\mu\nu}-\bar\Lambda\,f_{\mu\nu} = 0$, so that the background spacetime is some Einstein spacetime --- for example, Schwarzschild/ Kerr/ de~Sitter/ anti-de~Sitter, or 
even Milne spacetime, not necessarily Minkowski spacetime.
(Implicitly ignoring any background cosmological constant, as in~\cite{Volkov:2011an, Comelli:2011zm}, leads to
a Ricci-flat space for finite $\mathcal{T}_{\mu\nu}$.)
However, following the philosophy of massive gravity, the background metric should not be constrained by the limiting procedure, but 
externally specified
by the theory itself. Thus, if one wants to recover massive gravity as a limit of bimetric gravity, then it seems inconsistent 
to consider a limit which (unnecessarily) fixes the background metric.

\subsection{The non-dynamical background limit}\label{sec31}

If we consider the action of bimetric gravity given by equation~(\ref{actionbg}), it is easy to see that one can 
recover the action of massive gravity, equation~(\ref{actionmg}), simply by simultaneously taking the limits 
$\kappa\rightarrow0$ and $\epsilon\to 0$. 
Furthermore, the consideration of the limit $\kappa\rightarrow0$ will imply some constraints
on the kind of interaction between both gravitational sectors in
bimetric gravity, leading to one that must be also compatible with massive gravity to recover this theory, 
while not fixing the background metric as in references~\cite{Volkov:2011an,Comelli:2011zm}, at least in principle.
For consistency it can be checked that the 
same constraints coming from the consideration of this limit can be obtained by taking directly the variation of
the action of massive gravity (\ref{actionmg}) with respect to the background metric $f_{\mu\nu}$.
Therefore, if one were interested in recovering
the predictions of massive gravity in a certain limit of bigravity theory, then it would be natural to consider 
such a limit, which corresponds to the limit of vanishing kinetic term (that is the limit of a vanishing
effective Planck mass for one of the metrics).

Considering $\kappa\rightarrow0$ and $\epsilon\to0$ in the background equation of motion~(\ref{motionf}), and using the 
definition of $\mathcal{T}_{\mu\nu}$ embodied in equation~(\ref{Tf}), the following perhaps unexpected constraint can be obtained:
\begin{equation}\label{tau0}
 \tau^{\mu}{}_{\nu}=0.
\end{equation}
Moreover, taking into account equation~(\ref{T}), one can  note that equations~(\ref{tau0}) imply that the
effective energy-momentum tensor appearing in the equations of motion for the dynamical metric, equation~(\ref{motiong}),
can be written as
\begin{equation}\label{TL}
T^{({\rm eff})\mu}{}_{\nu}=-\delta^{\mu}{}_{\nu}\; L_\mathrm{int}.
\end{equation}
Thus, the constraint (\ref{Bianchi}), ultimately coming from the contracted Bianchi identity,  implies
\begin{equation}\label{BianchiL}
 \partial_{\lambda}L_\mathrm{int}=0.
\end{equation}
Therefore, from equations~(\ref{TL}) and (\ref{BianchiL}) we can conclude that the modification in the equations of motion 
(\ref{motiong}) due to a putative non zero-mass for the graviton, considered as
a limit of a theory with two dynamical metrics, is equivalent from the point of view of the physical foreground metric to simply introducing a cosmological constant, one which can at least in principle be either positive
or negative. That is
\begin{equation}
T^{({\rm eff})\mu}{}_{\nu}=-\delta^{\mu}{}_{\nu}\;\Lambda_{{\rm eff}}.
\end{equation}
On the other hand, although in the next section we will focus our attention on theories
of massive gravity
with a spherically symmetric background metric, 
it should be pointed out that we have obtained no restriction about the curvature or 
dynamics of the background metric. In particular,
there is no requirement for this metric to be Riemann flat (or Ricci-flat, or even an Einstein spacetime).
It could be argued that one can choose to consider massive gravity with an Einstein background metric, and one can in fact do it.
Nevertheless, that cannot be a requirement coming from seeing this theory as a particular limit of bimetric gravity, because in that 
case the adoption of that particular point of view would change the philosophy of the theory itself.

Finally, it should be noted that up to now, we have not assumed any particular form for the interaction term (only that it depends on
the metrics through $[\gamma^n]$), neither have we assumed any
symmetry for the metrics. Therefore, these results are completely general when studying the solutions for any massive gravity 
theory of that class as a limit of bimetric gravity.  Thus, one can already conclude that there are fewer solutions that are solutions
simultaneously of massive gravity and of bimetric gravity in the limit $\kappa\rightarrow0$ and $\epsilon\to 0$,
than of massive gravity alone. This is because the former solutions must be also solutions of massive gravity and in addition fulfill
some extra constraints, namely equations~(\ref{tau0}), (\ref{TL}) and (\ref{BianchiL}). Moreover, those solutions, if any, 
would be equivalent to simply
considering a foreground cosmological constant.

\subsection{Ghost-free cosmologies}\label{sec32}

We now focus our attention on the ghost-free case, which corresponds to the specific interaction term given by 
equation~(\ref{int}),
and study solutions of massive gravity which are continuous in the parameter space (that is solutions of both massive gravity 
and bimetric gravity in the non-dynamical background limit).
As we have concluded that those solutions mimic the effect of a cosmological constant in the foreground space, it is of particular
interest to consider cosmological solutions. Thus, we assume a spherically symmetric situation, 
where both metrics fulfill this symmetry.
Both for massive gravity and for bigravity theories
these metrics can be written in general as~\cite{Volkov:2011an}
\begin{equation}\label{metricg}
 g_{\mu\nu}\,dx^{\mu}\,dx^{\nu}=S^2\,dt^2-N^2\,dr^2-R^2\,d\Omega_{(2)}^2,
\end{equation}
and
\begin{equation}\label{metricf1}
 f_{\mu\nu}\,dx^{\mu}\,dx^{\nu}=\left(A dt + C dr\right)^2-\left(B dr - {SC\over N} dt \right)^2-U^2\,d\Omega_{(2)}^2,
\end{equation}
where all the metric coefficients are functions of $t$ and $r$. 

As is well known~\cite{Salam:1976as,Isham:1977rj,Blas:2005yk} (see also reference~\cite{Volkov:2011an} for the ghost-free theory),
in bimetric gravity the requirement of $T^{({\rm eff})}{}_{0r}=0$ (or, equivalently, $\tau^{0}{}_{r}=0$), 
which is implied by the consideration of a spherically 
symmetric scenario, leads in general to two classes of solutions: (i) those in which both metrics can be written in a diagonal way using the same
coordinate patch, and (ii) those with metrics which are not commonly diagonal. Nevertheless, when one considers 
solutions of bimetric
gravity with a vanishing kinetic term, such solutions must fulfill constraints (\ref{tau0}) instead of the 
equations of motion of the background metric (\ref{motionf}),
leading $\mathcal{G}^{\mu}{}_{\nu}$ unspecified and, therefore, being compatible with massive gravity.
This requirement leads to the conclusion that there is no Lorentzian-signature 
solution for equations~(\ref{tau0}) if 
$C\neq0$ (see \ref{appendix-c} for details) without conflicting 
with~\cite{Salam:1976as,Isham:1977rj,Blas:2005yk,Volkov:2011an} .
So we can simplify the background metric and take
\begin{equation}\label{metricf}
 f_{\mu\nu}\,dx^{\mu}\,dx^{\nu}=A^2 \,dt^2-B^2 \,dr^2-U^2\,d\Omega_{(2)}^2,
\end{equation}
Considering metrics~(\ref{metricg}) and (\ref{metricf}), it can be seen that there are two general solutions, 
and additionally two special-case solutions, to equations~(\ref{tau0}) and (\ref{BianchiL}) for any ghost-free massive gravity
theory whose interaction term can be described by (\ref{int}).


\subsubsection{First general solution.}
The first general solution we discuss is particularly interesting, since
it relates the two metrics in a very simple way --- a position-independent rescaling. The metrics satisfy
\begin{equation}\label{conformes}
 f_{\mu\nu}\,dx^{\mu}\,dx^{\nu}=D^2 \,g_{\mu\nu}\,dx^{\mu}\,dx^{\nu},
\end{equation}
where $D$ is a constant with $D\neq 1$ such that
\begin{equation}
 c_4=\frac{3+3c_3(D-1)}{(D-1)^2}.
\end{equation}
Explicitly
\begin{equation}
D = 1 + {3 c_3\over2 c_4} \pm \sqrt{ \left(1+ {3 c_3\over2 c_4}\right)^2- 1}.
\end{equation}
For this solution the graviton mass produces a term in the modified Einstein equations (\ref{motiong}), which can be described
by the effective stress tensor
\begin{equation}
T^{{\rm (eff)}\mu}{}_\nu=-[3+c_3(D-1)] \,(D-1)^2 \; \delta^\mu{}_\nu,
\end{equation}
which mimics the behavior of a positive cosmological constant if
$c_3(1-D)>3$, and a negative one otherwise. Therefore, these solutions can describe a universe which is accelerating as if this
acceleration would be originated by a cosmological constant. Moreover, this foreground universe would have exactly the same symmetry as the background
metric --- that is having a homogeneous and isotropic foreground universe is possible only if
the background metric is also FLRW with the same sign of spatial curvature.

We might have argued on general grounds for the existence of a solution of this type
for a generic interaction term and in the absence of any symmetry of the metrics
once taking into account equation~(\ref{tau0}). When calculating $\tau^\mu{}_\nu$,
 notice that it is proportional to 
$\partial L_{{\rm int}}/\partial \gamma^\nu{}_\mu$. Thus, one might guess the existence of some general 
(not necessarily unique) solution with  $\gamma^\mu{}_\nu=D\,\delta^\mu{}_\nu$,
where the value of $D$ is fixed by the theory. 
In fact, in view of the specific form of $L_{{\rm int}}$, one might reasonably infer a generic polynomial constraint on $D$.
(That is because $\tau \propto {\partial L_{{\rm int}}}/{\partial \gamma}\sim \sum_i {\partial e_i}/{\partial \gamma} \sim\sum_k p_k(\gamma)\, \gamma^k$;
so if $\gamma=D\,\I$ is a solution to $\tau=0$, then one must have
$\left[{\partial L_{{\rm int}}}/{\partial \gamma}\right]_{\gamma=D\,\I}=P(D)\,\I=0$ for some polynomial $P(D)$.)
It must be pointed out that Blas, Deffayet, and Garriga already claimed for the existence of solutions of the kind 
considered in equation~(\ref{conformes})
in bigravity
for an arbitrary interaction term between the metrics, where $D$ would be determined by the particular interaction term \cite{Blas:2005yk}.
The novelty in our study resides in the fact that we have concluded that
those solutions are solutions of massive gravity which are continuous in the parameter space.  
Moreover, for the particular interaction 
Lagrangian considered in this paper, solution~(\ref{conformes}) is not the only solution implying an Einstein 
manifold $g$, as it was the case for the particular theory considered in reference~\cite{Blas:2005yk} (see 
also reference~\cite{Aragone:1972fn} for a particular model),
because any solution of bimetric gravity with $\tau^{\mu}{}_{\nu}\propto\delta^\mu{}_{\nu}$ 
would lead to a foreground metric of that kind.

\subsubsection{Second general solution.}
The second general solution corresponds to the metric
\begin{equation}
\label{second}
  f_{\mu\nu}\,dx^{\mu}\,dx^{\nu}=\bar{D}^2\,S^2(r,t)\,dt^2-\bar{D}^2\,N^2(r,t)\,dr^2-D^2\,R^2(r,t)\,d\Omega_{(2)}^2,
\end{equation}
where $D$ and $\bar D$ are two separate constants satisfying
\begin{equation}
 \bar{D}=\frac{-D+c_3(D-1)+2}{c_3(D-1)+1},
\end{equation}
and
\begin{equation}
 c_4=\frac{-1-c_3+c_3^2(D-1)^2-c_3D}{(D-1)^2}.
\end{equation}
In this case, the extra term in the equations of motion~(\ref{motiong}) behaves as a positive contribution to the 
cosmological constant if $c_3(D-1)>-1$, leading to a foreground universe with accelerating expansion.

This solution is similar to the previous one, with the only difference that the $t$--$r$ sector of both metrics is conformally
related through one constant whereas the angular sectors, in which we imposed the symmetry, are related through another constant.
This fact would not lead to great differences, at least in principle.


\subsubsection{Two special case  solutions.}
There are two additional special-case solutions
that hold only for a specific relation between the parameters $c_3$ and $c_4$.
In particular, we need to impose that $c_4=-3/4\,c_3^2$.
Both solutions lead to an effect in the modified Einstein equations~(\ref{motiong}) equivalent to a negative 
contribution to the cosmological constant.
For the first special case solution, the background metric can be written as 
\begin{equation}\label{fB1}
  f_{\mu\nu}\,dx^{\mu}\,dx^{\nu}=D^2\,S^2(r,t)\,dt^2-B^2(r,t)\,dr^2-D^2\,R^2(r,t)\,d\Omega_{(2)}^2,
\end{equation}
whereas for the second one
\begin{equation}\label{fB2}
  f_{\mu\nu}\,dx^{\mu}\,dx^{\nu}=A^2(r,t)\,dt^2-D^2\,N^2(r,t)\,dr^2-D^2\,R^2(r,t)\,d\Omega_{(2)}^2.
\end{equation}
Although these solutions are compatible only with a particular sub-class of models, they now allow us to have
two possibly different cosmological metrics in the background and foreground.

Regarding the first kind of solutions, where the foreground and background metric are given by equations~(\ref{metricg}) and
(\ref{fB1}), respectively, let assume that there are solutions with a foreground metric of the FLRW-kind. 
That is
\begin{equation}\label{g1a}
  g_{\mu\nu}\,dx^{\mu}\,dx^{\nu}=a^2(t)\, dt^2-\frac{a^2(t)}{1-kr^2}dr^2-a^2(t)r^2d\Omega_{(2)}^2,
\end{equation}
with $t$ being the so-called FLRW \emph{conformal time} coordinate (it is not the more usually occurring 
FLRW proper time coordinate)
and the scale factor $a(t)$ fulfilling the equation of motion~(\ref{motiong}), which, for this symmetry and this coordinate system, 
can be expressed as
\begin{equation}
 3\,\frac{\dot{a}^2+ka^2}{a^4}=\Lambda-m^2(D-1)^2+8\pi G\rho_{m}.
\end{equation}
Thus, although the effect of the graviton mass is equivalent to a negative contribution to the cosmological constant,
the solution is accelerating if $\Lambda_{{\rm eff}}>0$, with $\Lambda_{{\rm eff}}=\Lambda-m^2(D-1)^2$.
The foreground spacetime (\ref{g1a}) is a solution of massive gravity continuous in the parameter space, if we consider
a theory with a background metric which can be written as (see \ref{appendix-c})
\begin{equation}\label{Blibre}
  f_{\mu\nu}\,dx^{\mu}\,dx^{\nu}=D^2\,a^2(t)\,dt^2-B^2(r,t)\,dr^2-D^2\,a^2(t)\,r^2\,d\Omega_{(2)}^2.
\end{equation}
As the function $B(r,t)$ is not constrained by our analysis, it can take any form, not necessarily
related with $a(t)$. Thus, all the massive gravity theories defined with a background metric of the form given by 
equation~(\ref{Blibre})
have FLRW solutions which are continuous in the parameter space.
One interesting example would be to consider $B^2(r,t)=D^2a^2(t)/(1-\bar{k}r^2)$, in
such a way that the background metric now has different spatial curvature as the physical metric.
One could also consider more exotic cosmologies compatible with metric~(\ref{Blibre}).

We can also reverse the logic of the problem and take a massive gravity theory defined with an isotropic and homogeneous
background metric. That is
\begin{equation}\label{f1a}
  f_{\mu\nu}\,dx^{\mu}\,dx^{\nu}=a^2(t)\,dt^2-\frac{a^2(t)}{1-kr^2}\,dr^2-a^2(t)\,r^2\,d\Omega_{(2)}^2,
\end{equation}
which includes the particular case of massive gravity theories with a de Sitter background metric.
In this case, requiring the fulfillment of the constraints~(\ref{tau0}) (and (\ref{BianchiL})), it can be seen that
the foreground metric can be written as
\begin{equation}
  g_{\mu\nu}\,dx^{\mu}\,dx^{\nu}=\frac{a^2(t)}{D^2}\,dt^2-N^2(r,t)\,dr^2-\frac{a^2(t)\,r^2}{D^2}\,d\Omega_{(2)}^2.
\end{equation}
Note that the background scale factor $a(t)$ does not now fulfill any Friedmann-like equation, since the background metric is not
constrained but externally specified by the theory itself.
We can obtain various different kinds of physical cosmologies,
being described by the foreground metric, by changing the function $N(r,t)$. As already mentioned, one particular solution
belonging to this class would be that assuming that both spaces are FLRW, obtained by fixing $N(r,t)=a^2(t)/[D^2(1-\bar{k}r^2)]$.
Nevertheless, it is probably more interesting to consider the solutions implying an anisotropic expanding universe 
for a massive gravity
theory defined by using a FLRW background metric, which could even have the maximal symmetry being a de Sitter space. 
That would be the case of any solution obtained by choosing $N(r,t)=b^2(t)/[D^2(1-\bar{k}r^2)]$, with $b(t)\neq a(t)$.

\vspace{2mm}
On the other hand, those solutions in which the background metric takes the form given by metric~(\ref{fB2}) have 
similar characteristics as those 
with (\ref{fB1}). The only difference is that in this case one can arbitrarily choose the function appearing in 
the temporal component of
the metric. That is, for a theory with a FLRW background metric~(\ref{f1a}), one has
\begin{equation}
  g_{\mu\nu}\,dx^{\mu}\,dx^{\nu}=S^2(r,t)\,dt^2-\frac{a^2(t)}{D^2(1-kr^2)}\,dr^2-\frac{a^2(t)\,r^2}{D^2}\,d\Omega_{(2)}^2,
\end{equation}
which again can be fixed to be homogeneous and isotropic but it also allows the consideration of more exotic situations.

Finally, it must be pointed out that the solutions presented here are, on one hand, more general that those already studied 
in the literature (for instance they can describe anisotropic cosmologies in one sector) and, on the other hand, more restrictive than
other solutions, as they are only solutions for a particular model 
(with the parameters appearing in the Lagrangian fulfilling a particular relation).

\subsection{Limit procedure without vanishing background matter}\label{sec33}

Let us briefly consider a slightly different limit which retains 
some effects of the background matter, that is  $\kappa\to 0$ with $\epsilon$ nonzero and fixed. 
In this case the action becomes
\begin{eqnarray}
S&=&-\frac{1}{16\pi G}\int d^4x\sqrt{-g} \left\{R(g) + 2\,\Lambda-2\,m^2 L_\mathrm{int}\right\} +S_{({\rm m})} +\epsilon\,\bar S_{({\rm m})},
\end{eqnarray}
with dependence on the background metric $f_{\mu\nu}$ both in the obvious place $L_\mathrm{int}(f,g)$ and in the action $\epsilon\,\bar S_{({\rm m})}$ for the background matter fields, which now  have only an extremely indirect influence on the foreground physical sector. 
In this case the equation of motion for the background metric $f_{\mu\nu}$ becomes a purely algebraic one
\begin{equation}\label{motionf2}
\mathcal{T}^{\mu}{}_{\nu} = - {\epsilon\,8\pi G\over m^2} \,\bar T^{({\rm m})\mu}{}_{\nu},
\end{equation}
in terms of the background matter. One still has \emph{some} equation constraining the background metric,  and there will be some 
constraint intertwining the background and foreground metrics, and this is basically unavoidable in any reasonable limit of bimetric 
gravity that is based on tuning the parameters in the bigravity action to specific values.
Nevertheless, it should be noted that one would not be able to directly detect the existence of any matter content in the background.
Therefore, although the theory resulting when considering the limit $\kappa\to 0$ (with $\epsilon$ nonzero and fixed) has a very different
motivation that massive gravity, it seems that we could not distinguish between them by measuring their physical ``foreground'' consequences. In other 
words, we cannot think in any physical prediction which would be affected by considering that $f_{\mu\nu}$ is given by the theory, or
constrained by some matter invisible to us.

\section{Summary and Discussion}\label{cuatro}

In this work we have explored the cosmological solutions for a massive gravity theory when viewed as a limit of taking a vanishing kinetic term 
in a bimetric theory, paying particular attention in the way this limit is taken. We have used a vierbein formulation based on that 
developed by Volkov in reference~\cite{Volkov:2011an} 
(see also references~\cite{Nibbelink:2006sz, Chamseddine:2011bu, Chamseddine:2011mu, Hinterbichler:2012cn}), 
a formalism that proved to be very powerful in the treatment of this type of calculation.

A first step is to realize that the solutions of massive gravity, taken as the limit of a bigravity theory, are in general
more constrained than the general solutions of massive gravity.
That is because in massive gravity there is only one set of equations, whereas if one considers this theory as a particular limit of
bimetric gravity, then additional constraints must be taken into account.
Thus, one cannot recover complete continuity in the physical 
predictions of the theory, since the solutions continuous in the parameter space of the theory are not the complete class
of solutions.

We have argued that massive gravity can be recovered from bimetric gravity by suitably taking the limit for 
$\kappa \rightarrow 0$ in the action describing the latter.
We have shown that this limit implies that the modification to the equation of
motion introduced by a non-zero mass of the graviton is equivalent to introducing  an extra contribution to the 
cosmological constant which, at this stage, can be 
positive or negative. An important fact to stress is that no restriction on curvature or dynamics of the background metric has been imposed.

In particular, we have focused on ghost-free cosmologies with both metrics, foreground and background, being spherically symmetric.
In the first place, we have shown that there are no non commonly diagonal solutions continuous in the parameter space.
In the second place, we have found two kinds of solutions for any ghost-free theory.
The first kind of solutions implies that the two metrics are proportional to each other, with the graviton mass interaction Lagrangian producing 
an extra contribution to the cosmological 
constant that can be either positive or negative, depending on some relation between the parameters of the theory.
The net result is that this solution 
describes an accelerating or decelerating universe which has the same symmetry as the background metric. 
The second kind of solutions implies a more complicated relation between the two metrics, 
but we still obtain a positive cosmological constant for certain values of $c_3$.
Moreover, in the third place, we have also seen that two more kind of solutions can be obtained by considering a very specific 
relation between the parameters
$c_3$ and $c_4$, and that these solutions are equivalent to inducing an extra negative contribution to the cosmological constant 
in the modified Einstein theory. These solutions 
allow us to consider different cosmological metrics, for the background and foreground, related by some unconstrained arbitrary functions.
In the first kind of solutions, when considering a spherically symmetric background metric and a FLRW physical metric, with a scale factor
fulfilling the Einstein equations, we can obtain a different spatial curvature for the two metrics, by particularly tuning 
the function appearing
in the radial component of the metric.
If instead we consider the background metric to be FLRW, the (background) scale factor is now not forced to fulfill any Friedmann equations 
since this metric does not have a dynamics. 
A particular interesting case is to consider a massive gravity theory with an isotropic and homogeneous background metric which can lead to
anisotropic expansion of the foreground space.
On the other hand, in the second kind of solutions, we can obtain different cosmologies by 
changing the function in the temporal component of the metric.

It must be emphasized that the solutions obtained in this paper are solutions of massive gravity continuous in the 
parameter space of this theory; that is, they are solutions simultaneously of both massive gravity and bimetric gravity in the limit
of a vanishing kinetic term.
Thus, these solutions are completely different from those of massive cosmologies presented in 
references~\cite{deRham:2011by, Chamseddine:2011bu, D'Amico:2011jj, Gumrukcuoglu:2011ew,vonStrauss:2011mq}, 
which are only solutions of massive
gravity, and from those bimetric cosmologies studied in references~\cite{Volkov:2011an, vonStrauss:2011mq, Comelli:2011zm}, which are not solutions 
when the limit of vanishing kinetic term is taken.
On the other hand, these solutions must also not be confused with the solutions obtained in 
references~\cite{Comelli:2011zm, Volkov:2011an}
where a different limit of bimetric gravity is taken to recover massive gravity, namely $\kappa\rightarrow\infty$.

On the other hand, in view of the mentioned results,
one could consider that the consideration of massive gravity as a limit of bimetric gravity leads
to a theory which is equivalent to general relativity. 
Nevertheless, if this conclusion remain unchanged
by the consideration of perturbative effects in the metric (note that the effective cosmological constant
appears when considering the particular solutions and not directly in the action), then
that would not be a problem of our treatment but of the theory of massive gravity itself. In that case one should
conclude that: either (i) massive gravity is continuous on
the parameter space only when it is equivalent to general relativity (with a particular value for the cosmological constant), or
(ii) $\epsilon$ cannot be zero and the dynamics of our universe is affected by some background invisible matter, 
or (iii) massive gravity
cannot be seen as a particular limit of bimetric gravity.
We have briefly considered option (ii) in this paper, but there is a rich phenomenology of models that might be 
explored in this case. It could be interesting to consider whether the kinds of models that
should be considered could be constrained by some property of the \emph{hidden matter} sector. For example, could it be reasonable  to require
that the classical energy conditions be satisfied in this sector? Should there be some relationship between the standard model of particle physics in
both sectors, or can they be completely independent? Or, could this hidden matter be the ``mirror matter'' speculated to restore complete symmetry
in the fundamental interactions~\cite{Okun:2006eb,Foot:1991bp}?

In any case, in many ways it seems that bimetric theories and massive gravity should be kept in conceptually separate compartments. 
While there is no doubt that the relevant actions are very closely related, as we have seen treating massive gravity as a 
limit of bimetric gravity is  fraught with considerable difficulty. In fact, whereas in massive gravity $L_{{\rm int}}$ simply gives mass
to the graviton, in bimetric (and multimetric) gravity this term is actually describing an interaction between
two (or more) metrics. Therefore, bimetric gravity could be considered as a model of a bi-universe, in which two different physical 
worlds are
coexisting, and interacting only through the gravitational sector (at a kinematical level the ``analogue spacetime'' programme
can be used to provide examples of multi-metric, though not multi-gravity, universes 
\cite{Jannes:2011em,Volovik,Barcelo:2001ah,Barcelo:2001cp}).
Moreover, such an interpretation would lead to a new
multiversal framework when considering multimetric gravitational theories \cite{Hinterbichler:2012cn}, which would be very different
to those previously considered in the literature, since it is not resulting from quantum gravity \cite{Everett}, 
inflationary theory \cite{Linde:1986fc}, string theory \cite{Susskind:2003kw}, or 
general relativity \cite{GonzalezDiaz:2007yb}, but it is a consequence of abandoning Einstein's 
theory. This type of multiverse is also rather different from the more exotic multiverse concept developed 
in references~\cite{comic1,comic2}.

\ack

PMM acknowledges financial support from the Spanish Ministry of
Education through a FECYT grant, via
the postdoctoral mobility contract EX2010-0854.
VB acknowledges support by a Victoria University PhD scholarship.
MV acknowledges support via the Marsden Fund and via a James Cook Fellowship, both administered by the Royal Society of New Zealand.

\appendix
\section{Some identities}\label{appendix-a}

In this appendix we collect some purely algebraic results.
For a general $n\times n$ matrix $X$ the symmetric polynomials $e_i(X)$ are defined by
\begin{equation}
\sum_{i=0}^n \lambda^i \; e_i(X) = \det(\I+\lambda X). 
\end{equation}
Thus, the symmetric polynomials can be recovered iteratively from Newton's identity
\begin{equation}\label{Newton}
 e_i(X) =\frac{1}{i} \sum_{j=1}^i (-1)^{j-1}  \; e_{i-j}(X) \; \tr[X^j],
\end{equation}
taking into account $e_0(x)=1$.
Since
\begin{equation}
\det(\I+\lambda X^{-1}) = {\det(X+\lambda \I)\over\det(X)} = {\lambda^n\det( \I+\lambda^{-1} X )\over\det(X)}, 
\end{equation}
we have
\begin{equation}
e_i(X^{-1}) = {e_{n-i}(X) \over\det(X)},
\end{equation}
which is a purely algebraic result which allows us to rewrite the 
interaction term in bimetric gravity in various useful ways~\cite{Hassan:2011zd}
(see \ref{appendix-b} for more details).
Furthermore, since
\begin{eqnarray}
\fl\sum_{i=0}^n \; \lambda^i e_i(\I+\epsilon X) &=& \det((1+\lambda) \I+\epsilon \lambda X)
= \sum_{i=0}^n (1+\lambda)^{n-i} (\epsilon\lambda)^i \; e_i(X)\\
&=&  \sum_{i=0}^n  \sum_{j=0}^{n-i} {n-i\choose j} \lambda^j (\epsilon\lambda)^i \; e_i(X)
=  \sum_{i=0}^n  \sum_{k=i}^{n} {n-i\choose n-k} \lambda^k \epsilon^i \; e_i(X)\\
&=&  \sum_{k=0}^n  \sum_{i=0}^{k} {n-i\choose n-k} \lambda^k \epsilon^i \; e_i(X)
= \sum_{i=0}^n  \lambda^i  \sum_{k=0}^{i} {n-k\choose n-i} \epsilon^k \; e_k(X),
\end{eqnarray}
we have the ``shifting theorem''
\begin{equation}
e_i(\I+\epsilon X)  = \sum_{k=0}^{i} {n-k\choose n-i} \, \epsilon^k \; e_k(X) = 
\sum_{k=0}^{i} {n-k\choose i-k} \, \epsilon^k \; e_k(X).
\end{equation}

\section{The interaction term in bimetric gravity}\label{appendix-b}

In this paper we have used an expression for the graviton mass term, equation~(\ref{int}), which emphasized the fact 
that ghost-free massive 
gravity only includes three parameters more than general relativity. The consideration of such a  $L_{{\rm int}}$ in massive 
gravity leads to 
the same equations of motion as those that can be obtained using a formulation of the interaction term in terms of $\gamma$ 
as used in reference~\cite{Hassan:2011vm}. 
In this appendix we will show that this is also the case in bimetric gravity when one cannot throw away the term depending
only on the background metric. That is, both formulations
are equivalent also in bi\-metric gravity, at least when considering both a foreground and background cosmological constant. 

We note that the ghost-free foreground-background interaction terms can in all generality be written as
\begin{equation}\label{intbg}
\sqrt{-g} \; L_\mathrm{int} = \sqrt{-g} \; \sum_{i=0}^4 k_i \; e_i(\gamma) =  \sqrt{-f} \; \sum_{i=0}^4 k_{4-i} \; e_i(\gamma^{-1}).
\end{equation}
Here we have used the explicit algebraic symmetry between $e_i(X)$ and $e_{n-i}(X^{-1})$ to exhibit an explicit interchange symmetry 
between foreground and background geometries. (See reference~\cite{Hassan:2011zd}.) 
Furthermore, in view of the fact that $\gamma=\I-K$, the shifting theorem yields
\begin{eqnarray}
e_1(\gamma) &=& 4 - e_1(K);\nonumber \\
e_2(\gamma) &=& 6 - 3 e_1(K) + e_2(K);\nonumber \\
e_3(\gamma) &=& 4 - 3  e_1(K) + 2 e_2(K) - e_3(K);\nonumber \\
e_4(\gamma) &=& 1- e_1(K) + e_2(K)- e_3(K) + e_4(K).\nonumber
\end{eqnarray}
Consider the original ``minimal'' Lagrangian for generating a graviton mass~\cite{Hassan:2011hr}:
\begin{equation}
L_{\mathrm{old\,minimal}} = \tr\left(\sqrt{g^{-1} f}\right) - 3 = \tr(\gamma)-3 = e_1(\gamma)-3 = 1-e_1(K).
\end{equation}
Using the last of the shifting theorem equivalences from~\ref{appendix-a}, which relates $e_4(\gamma)$ with the polynomials in 
$K$, we see
\begin{equation}
L_\mathrm{old\,minimal} = e_4(\gamma) -  e_2(K)+ e_3(K) - e_4(K).
\end{equation}
When written in this way the $e_4(\gamma)$ term corresponds in bimetric gravity to a background cosmological constant, and in massive gravity to an irrelevant constant. The $e_2(K)$, $e_3(K)$, and $e_4(K)$ terms are manifestly quadratic, cubic, and quartic in $K$. So insofar as one is only interested in giving the graviton a mass the quantity 
\begin{equation}
L_\mathrm{new\,minimal} = -  e_2(K),
\end{equation}
does just as good a job (in fact arguably a better job) than the original minimal mass term.
This argument generalizes, we can use the shifting theorem to rewrite the general interaction term as
\begin{equation}
 L_\mathrm{int} = \sum_{i=0}^4 k_i \; e_i(\gamma) =  \sum_{i=0}^4 \tilde k_i \; e_i(K).
 \end{equation}
Explicitly separating off the first two terms and using again the last of the shifting theorem equivalences we see
\begin{equation}
 L_\mathrm{int} = \tilde k_0 + \tilde k_1 \left\{ 1-e_4(\gamma) +  e_2(K)- e_3(K) + e_4(K) \right\} +  \sum_{i=2}^4 \tilde k_i \; e_i(K).
\end{equation}
So we have
\begin{eqnarray}
\fl L_\mathrm{int} =  (\tilde k_0 + \tilde k_1) \, e_0(K) + (\tilde k_2+\tilde k_1) \, e_2(K) +  (\tilde k_3-\tilde k_1) \, e_3(K) 
 + (\tilde k_4+\tilde k_1) \, e_4(K)  \nonumber\\
  - \tilde k_1 \, e_4(\gamma).
\end{eqnarray}
This eliminates (or rather redistributes) the $e_1(K)$ term.
For current purposes it is now useful to split off the top and bottom terms (corresponding to foreground and background cosmological constants) and deal with them separately. 
We have explicitly absorbed them into the bigravity kinetic terms. After factoring out an explicit $m^2$, this finally leaves us with the 
three-term interaction Lagrangian we have used in the paper, equation~(\ref{int}). This is
\begin{equation}\label{intmg}
L_\mathrm{int} =  e_2(K) -  c_3 \; e_3(K)   - c_4 \; e_4(K).
\end{equation}
Thus, we have shown explicitly that the interaction term~(\ref{intmg}) describes the same bimetric gravity 
theory as that considered in~(\ref{intbg})
when one takes into account a cosmological constant for each metric.

On the other hand, there is still a hidden ``symmetry'' between foreground and background. 
We can define an equivalent $K_f$ for the background metric with respect to the foreground metric via
\begin{equation}
K_f = \I - \gamma^{-1} = \I + (\I-K)^{-1} = - K (\I-K)^{-1},
\end{equation}
and equivalently
\begin{equation}
K = - K_f (\I - K_f)^{-1}.
\end{equation}
Note that
\begin{equation}
e_2(K) = e_2(K_f) + \mathcal{O}(K_f^3).
\end{equation}
Therefore, as could have been suspected from equation~(\ref{intbg}), one can equivalently consider that the interaction term is giving mass
either to the graviton related with $f_{\mu\nu}$ or to that of $g_{\mu\nu}$. In fact, the ghost-free bimetric gravity is giving us 7 degrees
of freedom (14 considering also the conjugate momenta) to distribute between both gravitons (see reference~\cite{Hassan:2011vm}), 
without any particular 
preference as to which. Here resides the great conceptual difference between massive gravity and bimetric gravity, since whereas
in the first theory $L_{{\rm int}}$ is a term whose purpose is giving mass to the graviton (the only graviton present in this theory), in
bimetric gravity the non-existence of a preferred metric leads one to interpret $L_{{\rm int}}$ merely as an interaction term. In fact, one
could even think of some kind of {\it democratic principle} for bimetric theories, by interpreting the degrees of freedom to be
distributed in such a way that we have two massless gravitons, and an interaction between the two metrics mediated by one vectorial and 
one scalar degree of freedom.

\section{Spherically symmetric solutions}\label{appendix-c}

Taking into account the spherically symmetric
metrics (\ref{metricg}) and (\ref{metricf1}) in equation~(\ref{tau}),
the non-vanishing components of $\tau^\mu{}_\nu$ can be written as~\cite{Volkov:2011an}
\begin{eqnarray}
 \fl \tau^0{}_0=\frac{A\,B}{S\,N}+\frac{C^2}{N^2}-\frac{3\,A}{S}+\frac{2\,A\,U}{S\,R}+c_3\left(1-\frac{U}{R}\right)
 \left(\frac{3\,A}{S}-\frac{2\,A\,B}{S\,N}-\frac{2\,C^2}{N^2}-\frac{A\,U}{S\,R}\right)
 \nonumber\\
  +c_4\left(1-\frac{U}{R}\right)^2\left(\frac{A}{S}-\frac{A\,B}{S\,N}-\frac{C^2}{N^2}\right),
\end{eqnarray}
\begin{eqnarray}
\fl \tau^r{}_r=\frac{A\,B}{S\,N}+\frac{C^2}{N^2}-\frac{3\,B}{N}+\frac{2\,B\,U}{N\,R}+c_3\left(1-\frac{U}{R}\right)
 \left(\frac{3\,B}{N}-\frac{2\,A\,B}{S\,N}-\frac{2\,C^2}{N^2}-\frac{B\,U}{N\,R}\right)
 \nonumber\\
 +c_4\left(1-\frac{U}{R}\right)^2\left(\frac{B}{N}-\frac{A\,B}{S\,N}-\frac{C^2}{N^2}\right),
\end{eqnarray}
\begin{eqnarray}
\fl \tau^\theta{}_\theta=\tau^\phi{}_\phi=
 c_3\frac{U}{R}\left(3-\frac{2\,B}{N}-\frac{2\,U}{R}+\frac{B\,U}{N\,R}-\frac{2\,A}{S}+\frac{A\,U}{S\,R}+\frac{A\,B}{S\,N}+\frac{C^2}{N^2}\right)
 \nonumber\\
 +\frac{U}{R}\left(\frac{A}{S}+\frac{B}{N}-3+\frac{U}{R}\right)
 \nonumber\\+c_4\frac{U}{R}\left(1-\frac{U}{R}\right)\left(1-\frac{A}{S}-\frac{B}{N}+\frac{A\,B}{S\,N}+\frac{C^2}{N^2}\right),
\qquad
\end{eqnarray}
and finally
\begin{eqnarray}
 \tau^0{}_r=\frac{C}{S}\left[-3+2\frac{U}{R}+c_3\left(3-\frac{U}{R}\right)\left(1-\frac{U}{R}\right)+c_4\left(1-\frac{U}{R}\right)^2\right].
\end{eqnarray}
In particular this implies
\begin{equation}\label{resta}
 \tau^r{}_r-\tau^0{}_0=\frac{B\,S-A\,N}{C\,N}\;\tau^0{}_r,
\end{equation}
which greatly simplifies some calculations.
We must find solutions for all these components being set equal to zero. Let us start by requiring
that $\tau^0{}_r=0$. This can be obtained in either of two ways:
\begin{enumerate}
\item $C\neq0$, with $U(r,t)= D\; R(r,t)$, and an appropriate constraint on $D$.
\item $C=0$. 
\end{enumerate}

\subsection{Non-diagonal background metric}

If we wish to consider a non-diagonal background metric, then we must require $ U(r,t)=D\cdot R(r,t)$
in order to have  $\tau^0{}_r=0$, where $D$ is a constant such that
\begin{equation}
c_4=\frac{3-2D-c_3\left(3-4D+D^2\right)}{(D-1)^2},
\end{equation}
and, of course, $D\neq1$. Substituting this into $\tau^r{}_r-\tau^{\theta}{}_{\theta}$ we find that
$\tau^r{}_r-\tau^{\theta}{}_{\theta}$ vanishes when 
\begin{equation}
c_3=\frac{D-2}{D-1},
\end{equation}
implying
\begin{equation}
 c_4=-\frac{3-3D+D^2}{(D-1)^2}.
\end{equation}
The components of  $\tau^\mu{}_\nu$ then reduce to
\begin{equation}
 \tau^\mu{}_\nu=D(D-1)\left(\frac{C^2}{N^2}+\frac{A\,B}{N\,S}\right)\;\delta^\mu{}_\nu.
\end{equation}
Thus, one must require $A=-C^2S/(N\,B)$ to have a non-trivial situation.
The interaction term now reads
\begin{equation}
 L_{{\rm int}}=D(D-1)\left(\frac{A\,B}{N\,S}+\frac{C^2}{N^2}-\frac{1}{D}\right) = - (D-1).
\end{equation}
Replacing the relation  $A=-C^2S/(N\,B)$ in equation~(\ref{metricf}) for the background metric, one has
\begin{eqnarray}\label{metricfups}
 f_{\mu\nu}\,dx^{\mu}\,dx^{\nu}&=&\frac{C^2S^2}{N^2}\,\left(\frac{C^2}{B^2}-1\right)\,dt^2+2\frac{C\,S\,B}{N}\,\left(-\frac{C^2}{B^2}+1\right)\,dt\,dr
 \nonumber\\
&&- B^2\,\left(-\frac{C^2}{B^2}+1\right)\,dr^2-D^2\,R^2\,d\Omega_{(2)}^2,
\end{eqnarray}
which is non-Lorentzian. In fact this metric can be written as
\begin{eqnarray}\label{metricfups2}
 f_{\mu\nu}\,dx^{\mu}\,dx^{\nu}&=&\left(\frac{C^2}{B^2}-1\right)\left( {CS\over N} \,dt - B dr \right)^2 -D^2\,R^2\,d\Omega_{(2)}^2,
\end{eqnarray}
which manifestly has unphysical null signature $(0,-\mathrm{sign}[B^2-C^2],-1,-1)$. Therefore there is no physical solution of this kind.

\subsection{Diagonal background metric}

We now set $C=0$. 
In order to have $\tau^r{}_r-\tau^0{}_0=0$ we have two possibilities, either $ U(r,t)=D\cdot R(r,t)$ with $D\neq1$ and such that
\begin{equation}
c_4=\frac{3-2D-c_3\left(3-4D+D^2\right)}{(D-1)^2},
\end{equation}
as before (``case I''), or $B\,S=A\,N$ (``case II'').

 \paragraph{Case I:} \underline{Solutions for particular models.}
If we consider the first case, $U=D\,R$, then in order to have also $\tau^r_r-\tau^\theta_\theta=0$, there are three options:
\begin{enumerate}
\item $c_3$ takes the same value as in the previous subsection (when considering a non-diagonal background metric): this implies the same consequences. The background metric is non-Lorentzian now specifically with unphysical null signature $(0,-1,-1,-1)$.

\item $A=D\,S$: In this case we have
\begin{equation}
 \tau^\mu{}_\nu=\frac{B\,(2+c_3(D-1))(D-1)D}{N}\;\delta^\mu{}_\nu,
\end{equation}
thus, the theory should have
\begin{equation}\label{c1a}
 c_3=-\frac{2}{D-1},\,\,{\rm and}\,\,\,  c_4=-\frac{3}{(D-1)^2},
\end{equation}
to have solutions. The Lagrangian is $L_\mathrm{int}=(D-1)^2$, which fulfills the Bianchi-inspired constraint.

\item $B=D\,N$: Now
\begin{equation}
 \tau^\mu{}_\nu=\frac{A\,(2+c_3(D-1))(D-1)D}{S}\;\delta^\mu{}_\nu,
\end{equation}
which must vanish. Therefore, we have again $c_3$ and $c_4$ given by equation~(\ref{c1a}), and the same
$L_{{\rm int}}$ as in the previous case.
\end{enumerate}
\paragraph{Case II:} \underline{Solutions for all the ghost-free models.}
We now consider $A=B\,S/N$. It can be seen that there are two cases in which $\tau^r{}_r-\tau^\theta{}_\theta=0$. These are:
\begin{enumerate}
 \item $B=N\,U/R$: We then have
\begin{equation}
 \fl \tau^\mu{}_\nu=-\frac{(R-U)U[(3-3c_3-c_4)R^2+(3c_3+2c_4)R\,U-c_4U^2]}{R^4}\; \delta^\mu{}_\nu.
\end{equation}
These quantities vanish for $U=D\cdot R$, with $D$ such that
\begin{equation}
 c_4=\frac{3+3c_3(D-1)}{(D-1)^2},
\end{equation}
which implies $L_\mathrm{int}=(3+c_3(D-1))(D-1)^2$. Thus, we should require $D\neq1$ to have a (non-trivial) massive gravity theory.

 \item Consider
 \begin{equation}
B=N\;  \frac{3-3c_3-c_4+(2c_3+c_4-1)U/R}{1-2c_3-c_4+(c_3+c_4)U/R}.
\end{equation}
 In this case we have
 \begin{eqnarray}
\fl\tau^{\mu}{}_{\nu}=\frac{U}{R^{2}}\; \frac{[(3-3c_3-c_4)R-(1-2c_3-c_4)U]}{[(2 \,c_3+c_4-1)- U( c_3 + c_4) ^{2}} 
\nonumber \\
\fl\qquad\times[(c_3^2-c^3+c_4+1)R^2+(c_3-2c_3^2-2c_4)R\,U+(c_3^2+c_4)U^2] \;\;\delta^\mu{}_\nu,
\end{eqnarray}
which again vanishes if $U=D\cdot R$, but now we must have
 $D$ such that
\begin{equation}
 c_4=-\frac{c_3^2(D-1)^2+c_3(D-1)+1}{(D-1)^2}.
\end{equation}
The mass term now reads
\begin{equation}
 L_{{\rm int}}=-\frac{(D-1)^2}{c_3(D-1)+1}.
\end{equation}

It should be noted that there is yet one more formal solution for the equations $\tau^\mu{}_\nu=0$, 
but one which at the end of the day implies $B=0$. Thus, for that solution the background metric is again 
non-Lorentzian and unphysical, this time with null signature $(+1,0,-1,-1)$.
\end{enumerate}
This completes our consideration of the various explicit constraints on the background geometry arising from the equation 
$\tau=0$ in the case of spherical symmetry.

\end{document}